\documentclass[12pt]{article}
 \usepackage{graphicx}
 \usepackage{amssymb,amsmath}
\begin{document}
\begin{center}
{\Large\bf \boldmath Annihilating States in Close-Coupling Method
for Collisions between Hadronic and Ordinary Atoms}

\vspace*{6mm} {G.Ya. Korenman and S.N. Yudin}\\
 {\small \it D.V.Skobeltsyn Institute of Nuclear Physics,
 M.V. Lomonosov Moscow State University,\\
Moscow 119991,  Russia}

E-mail: korenman@nucl-th.sinp.msu.ru
\end{center}

\vspace*{6mm}

\begin{abstract}
Traditional close-coupling methods suppose an expansion of the
total wave function in terms of inner stationary states of
colliding subsystems. In the case of hadronic atoms, a similar
expansion has to involve, \emph{inter alia}, low angular momentum
states ($ns,\,np$) with large annihilation or nuclear absorbtion
widths. The life times $\tau_{nl}=\hbar/\Gamma_{nl}$ of these
states are small as compared with the collision time and mean time
between subsequent collisions, therefore the close-coupling
approach has to be modified for the similar problems.

In this paper we propose a generalization of the close-coupling
method with annihilating states included in the basis. The correct
asymptotic behaviour of the wave function in the annihilating
channels suppose that the annihilating states can not be presented
in the incoming channels whereas the corresponding components of
the wave function of relative motion in outgoing channels have to
damp out at large distances. The S-matrix of the transitions in
the subspace $\alpha$ of all other (non-annihilating) states is
not unitary, because the hamiltonian of the problem is
non-hermitian. The unitary defect $(1- \sum |S_{ji}|^2)$ gives the
cross section of induced annihilation for the initial state
$i\in\alpha$. The scheme of the numerical solving of the
generalized close-coupling equations is proposed. It includes the
calculations of two types of solution, from the origin to an
intermediate sewing point $R=a$ and from a large $R$ to the point
$a$.

\end{abstract}

\vspace*{6mm}

\section{Introduction}
Quantum close-coupling method is used in the many theoretical
problems of atomic and nuclear two-body collisions $A + B
\rightarrow A' + B'$. It supposes an expansion of the total wave
function $\Psi(\mathbf{R},\xi)$ of the system in terms of inner
wave functions $\phi_k(\xi)$ of colliding subsystems,
\begin{equation} \label{eq1}
\Psi(\mathbf{R},\xi)=\sum_k \phi_k(\xi) \psi_k(\mathbf{R}),
\end{equation}
where the coefficients $\psi_k(\mathbf{R})$ have a sense of wave
functions of the relative motion of two subsystems ($A$ and $B$)
in the corresponding channels. They satisfy to the system of
coupled equation with the standard boundary conditions at
$R\rightarrow\infty$ (incoming and outgoing waves in the open
channels, and damping waves in the closed channels). The expansion
\eqref{eq1} and the mentioned boundary conditions suppose that the
basis states $\phi_k(\xi)$ are stationary. However excited states
of real quantum systems (atoms, nuclei, \emph{etc.}) have finite
life times. Therefore the mentioned approach can be used if the
life times of the states involved in the consideration are large
as compared with the collision time, $ \tau_k\gg
\tau_{\mathrm{coll}}\sim R_0/v$, where $R_0$ is an interaction
radius and $v$ is the relative velocity of the colliding
subsystems. This condition, as a rule, is even not mentioned in
the papers that used the close-coupling methods, because it is
fulfilled in the most part of the problems of atomic and nuclear
collisions. However, there is, at least, one class of the problems
where the mentioned condition is strongly violated, namely, when
one of the colliding subsystems is an hadronic atom. In this case
the expansion \eqref{eq1} has to involve, \emph{inter alia}, low
angular-momentum states ($ns,\,np$) with large annihilation
(nuclear absorbtion) widths. The life times
$\tau_{nl}=\hbar/\Gamma_{nl}$ of these states are small as
compared with the collision time and mean time between subsequent
collisions, and therefore these states can not be presented among
possible incoming channels. They can be admixed to the total wave
function in the area of interaction $(R\lesssim R_0)$, but then
they disappear during collision. The corresponding components
$\psi_{nl}(\mathbf{R})$ of the wave function of relative motion in
outgoing channels have to damp out at large distances. The
parameter
 \[\nu_{nl}=\tau_{\mathrm{coll}}/\tau_{nl} \sim
 \Gamma_{nl} R_0/\hbar v \]
gives a criterion of the state stationarity during collisions. At
$\nu_{nl} \ll 1$ the non-stationarity during collision can be
neglected, whereas for the states with $\nu_{nl}\gtrsim 1$ the
non-stationarity should be taken into account. We will refer
similar short-lived states as \emph{annihilating} states.

The problems of very short-lived states in collisions are not
specific for the close-coupling method. They would arise in any
theoretical approach to the collisions of the similar systems.
However they are considered up to now only in the framework of
semiclassical approximations (see, e.g., \cite{ref1,ref2}) that
hold for "hot" hadronic atoms ($ka\gg 1$, or, for
$(p\bar{p})_{nl}+\mathrm{H}$ collisions, $E\gtrsim 1$ eV). To our
knowledge the single attempt to take into account nuclear
absorbtion in $ns$-states during collisions within the quantum
close-coupling method was done in the paper \cite{ref3} for the
$(\pi^- p)_{nl} +\mathrm{H}$ scattering. However the authors  use
an artificial assumption that the widths $\Gamma_{ns}$ are turned
off at some distance between two atoms $(R>R_0= 5a_0)$ that,
evidently, contradicts to the physical reality.

In this paper we propose a generalization of the close-coupling
method with short-lived (annihilating) states included in the
basis. The boundary conditions at $R\rightarrow\infty$ in the
channels related to the annihilating states are formulated so that
they have the correct quantum mechanical meaning in accordance
with the physical properties of these states.

\section{Close-coupling equations with short-lived states included in the basis}

Let us consider multichannel two-body collisions of hadronic atom
or ion $(\bar{h}Z)_{nl}$ with a neutral atom $A$
\begin{equation} \label{eq2}
(\bar{h}Z)_{nl} + A \rightarrow (\bar{h}Z)_{n'l'} + A,
\end{equation}
where $\bar{h}$ is a negative hadron ($\pi^-$, $K^-$, $\bar{p}$,
\emph{etc.}), $Z$ is a bar nucleus (e.g., $\mathrm{H}^+$ or
$\mathrm{He}^{2+}$). For the sake of simplicity we suppose that
(i) the hadronic atom does not contain electrons, therefore the
states of the $(\bar{h}Z)$ are classified by the principal quantum
number $n$ and the quantum numbers of orbital angular momentum $l$
and $m$, (ii) the energy $E$ of the collision is small ($E\ll
E_a$, where $E_a=27.2$ eV is the atomic unit of energy) so that
the channels with the excited atom $A^*$ are closed and can be
neglected, (iii) the principal quantum number $n$ is large enough,
therefore the orbital velocity of the hadron $v_n=Zv_a/n \ll v_e$,
where $v_a=e^2/\hbar$ is the atomic unit of velocity and $v_e\sim
Z_{eff} v_a$ is a characteristic electron velocity in the atom
$A$. The conditions on $E$ and $n$ mean that the heavy particles
(nuclei and hadron) in the problem under consideration are slow as
compared with electrons. Therefore electronic variables can be
separated out within adiabatic approximation. Then the sum of
Coulomb interactions between the hadronic atom $(\bar{h}Z)$ and
the many-electronic atom $A$ is replaced by the effective
three-body interaction $V(\mathbf{R,\, r})$ that depends on the
relative coordinates $\mathbf{R}$ of two colliding subsystems and
on the inner coordinates $\mathbf{r}$ of the hadronic atom. Total
effective hamiltonian of the three heavy particles $(\bar{h}-Z-A)$
can be presented as
 \begin{equation} \label{eq3}
  H = T(\mathbf{R}) + h(\mathbf{r}) + V(\mathbf{R,\, r}) ,
\end{equation}
where $T(\mathbf{R}) = (-1/2m)\nabla^2_{\mathbf{R}}$ is the
kinetic energy operator, $m$ is the reduced mass of the colliding
subsystems, and $h(\mathbf{r})$ is the inner hamiltonian of the
hadronic atom.

The hamiltonian of the hadronic atom could be, in general, very
complicated due to possible transmutations of the $(\bar{h}+Z)$
system into other hadronic systems. However for our aims it can be
presented as
\begin{equation} \label{eq4}
h =h_0 +U_{\mathrm{opt}}(r),
\end{equation}
where $h_0$ is the hamiltonian of hydrogen-like atom with the
reduced mass $\mu$ and with the nuclear charge $Z$, and
$U_{\mathrm{opt}}(r)$ is a short-range complex optical potential
with the absorptive imaginary part
$\mathrm{Im}U_{\mathrm{opt}}(r)\leq 0$. Let $E_{nl}$ and
$\phi_{nlm}(\mathbf{r})=R_{nl}(r)Y_{lm}(\Omega_r)$ be the
eigenvalues and the eigenfunctions of the hamiltonian \eqref{eq4},
corresponding to discrete levels of the $(\bar{h}Z)$ system,
\begin{equation} \label{eq5}
h \phi_{nlm}(\mathbf{r}) = E_{nl}\phi_{nlm}(\mathbf{r}).
\end{equation}
The hamiltonian $h$ is non-hermitian $(h^+\neq h)$ due to the
imaginary part of the optical potential. Therefore the eigenvalue
$E_{nl}$ in general case also can contain a negative imaginary
part,
\begin{equation}\label{eq6}
E_{nl}=E_{nl}^R - \mathrm{i}\Gamma_{nl}/2,
\end{equation}
where $\Gamma_{nl}$ is the annihilation or nuclear absorption
width of the level. The real part of the eigenvalue is usually
expressed as $E_{nl}^R=e_n - \epsilon_{nl}$, where $e_n = -\mu
Z^2/2n^2$ is the energy of the hydrogen-like system with the point
nuclear charge, and $(-\epsilon_{nl})$ is the energy shift of the
level due to the nuclear interaction and the finite distribution
of the nuclear charge. (We use the atomic system of units
$\hbar=e=m_e=1$, except as otherwise indicated explicitly.) Total
difference $(E_{nl} - e_n)$ is usually referred as the complex
energy shift $\Delta E_{nl}=-\epsilon_{nl} -
\mathrm{i}\Gamma_{nl}/2$.

The conjugated hamiltonian $h^+$ has the complex-conjugated
eigenvalues $\widetilde{E}_{nl}=E^*_{nl}$ and the eigenfunctions
$\widetilde{\phi}_{nlm}(\mathbf{r})=\widetilde{R}_{nl}(r)Y_{lm}(\Omega_r)$,
where $\widetilde{R}_{nl}(r)$ coincides with $R^*_{nl}(r)$. Two
sets of the eigenfunctions, $\phi_{nlm}(\mathbf{r})$ and
$\widetilde{\phi}_{nlm}(\mathbf{r})$, form jointly a bi-orthogonal
system,
\begin{equation} \label{eq7}
\langle \widetilde{\phi}_{nlm}|\phi_{n'l'm'} \rangle = \langle
\phi_{nlm}|\widetilde{\phi}_{n'l'm'} \rangle = \delta_{nn'}
\delta_{ll'}\delta_{mm'} .
\end{equation}

In order to introduce a close-coupling expansion of the total wave
function of the system $(\bar{h}Z)+A$, we define the basis states
\begin{equation} \label{eq8}
\Phi_j^{JM\pi}(\mathbf{r},\Omega_R)=
\left(\phi_{nl}(\mathbf{r})\otimes Y_L(\Omega_R)\right)_{JM} .
\end{equation}
Here $\Omega_R=(\theta,\phi)$ are polar and azimuthal angles of
the vector $\mathbf{R}$, $L$ is the quantum number of the relative
angular momentum of the colliding subsystems, and
$\left(\phi_{nl}\otimes Y_L\right)_{JM}$ means the vector coupling
of the states of two angular momenta ($l$ and $L$) into the total
angular momentum $J$. Index $j=\{n,l,L\}$ enumerates the basis
states with the definite quantum numbers of the total angular
momentum $J,M$ and of the parity $\pi=(-1)^{l+L}$. If
$\phi_{nlm}(\mathbf{r})\neq \widetilde{\phi}_{nlm}(\mathbf{r})$,
we have to introduce also the states
$\widetilde{\Phi}_j^{JM\pi}(\mathbf{r},\Omega_R)$ that are differ
from \eqref{eq8} only by substituting
$\widetilde{\phi}_{nl}(\mathbf{r})$ for $\phi_{nl}(\mathbf{r})$.
In this case we obtain again the bi-orthogonal system,
\begin{equation} \label{eq9}
\langle \widetilde{\Phi}_j^{JM\pi} | \Phi_i^{J'M'\pi'} \rangle =
\langle \Phi_i^{J'M'\pi'} |\widetilde{\Phi}_j^{JM\pi} \rangle =
\delta_{ij} \delta_{JJ'}\delta_{MM'}\delta_{\pi\pi'} .
\end{equation}
Total wave function of the $(\bar{h}Z)+A$ system in the state with
the total energy $E$, the total angular momentum $J$ and the
parity $\pi$ in the framework of the close-coupling approximation
can be written as
\begin{equation} \label{eq10}
 \Psi_i^{EJM\pi}(\mathbf{R,r}) =\sum_k \Phi_k^{JM\pi}(\mathbf{r},\Omega_R) \psi_{ki}^{EJ\pi}(R)/R,
 \end{equation}
where $i$ is the index of an incoming channel. The conserved
quantum numbers $(E,J,M,\pi)$ will be omitted in the further
equations the for the sake of simplicity.

Substituting the expansion \eqref{eq10} into the stationary
Schroedinger equation with the effective hamiltonian \eqref{eq3}
and taking into account the bi-orthogonality property \eqref{eq9},
we get the system of coupled-channel equations
\begin{equation} \label{eq11}
 \psi_{ji}^{\prime\prime}(R) +  \left[k_j^2- L_j(L_j+1)/R^2\right] \psi_{ji}(R)
= 2m\sum_k V_{jk}(R)\psi_{ki}(R),
\end{equation}
where $k_j^2=2m(E-E_{nl})$ and $ V_{jk}(R)= \langle
\widetilde{\Phi}_j |V(\mathbf{R,\, r})| \Phi_k \rangle $. The
system \eqref{eq11} looks very similar to that one commonly used
in the standard close-coupling methods. However in the problem
under consideration the wave numbers $k_j$ in some channels can
contain imaginary parts, therefore the conventional boundary
conditions at $R\rightarrow\infty$ (incoming + outgoing waves) in
these channels have no sense and should be modified. For
simplicity, we will not consider the closed channels with
$E_{nl}^R>E$. As for the behaviour of $\psi_{ji}(R)$ at
$R\rightarrow 0$, the standard requirement of the regular
solutions and the boundary conditions at the origin
$\psi_{ji}(0)=0$ remain valid.

Let us divide the total space of the channels included in the
basis into two subspaces, $\alpha$ and $\beta$. The subspace
$\alpha$ contains the channels that correspond to the stationary
states with the negligible widths $(\Gamma_{nl}\rightarrow 0)$
and, respectively, to the very small values of the parameter
$\nu_{nl}$, whereas the subspace $\beta$ contains annihilating
channels ($\nu_{nl}\gtrsim 1$). As a rule, $ns$-states of hadronic
atoms fall into the subspace $\beta$. In addition, depending on
the physical system and the kinetic energy under consideration,
other states with small angular momenta can also fall into this
subspace. The absorption widths $\Gamma_{nl}$ decrease drastically
with angular momentum, therefore the states with large $l$ always
belong to the subspace $\alpha$. The wave numbers $k_j$ for the
channels $j\in\alpha$ are real, whereas for the channels
$j\in\beta$ they are complex. In the latter case we define the
wave number so that $\mathrm{Im}k_j\geqslant 0$.

It is evident from the physical considerations that the channels
from the subspace $\beta$ can not be presented among the incoming
channels, because the hadronic system $(\bar{h}Z)$ in the
short-lived state will disappear before the arrival from a far
distance into the interaction area. Therefore the corresponding
components of the total wave function have to be zero everywhere
in $R\in[0,\infty)$:
\begin{equation} \label{eq12}
\psi_{ji}(R) \equiv 0 \text{ at } i\in \beta \text{ at any } j.
\end{equation}
(Recall that the second index in $\psi_{ji}(R)$ enumerates the
incoming channels.)

The remaining nonzero solutions $\psi_{ji}(R)$ correspond to the
ordinary incoming channels $(i\in\alpha)$. The components with
$j\in\alpha$ can contain the both incoming (at $j=i$) and outgoing
waves, whereas the components with $j\in\beta$ have to damp out at
the large distance. These latter can be presented in the form of
'outgoing' waves with the complex $k_j$ ($\mathrm{Im} k_j
\geqslant 0$). Hence the asymptotic boundary conditions at
$R\rightarrow\infty$ for the functions $\psi_{ji}(R)$ at
$i\in\alpha$ and any $j$ can be written as
 \begin{equation} \label{eq13}
\psi_{ji}(R) \rightarrow k_i^{-1/2} \exp[-\mathrm{i}(k_i
R-L_i\pi/2)] \delta_{ij} - k_j^{-1/2}\exp[\mathrm{i}(k_j
R-L_j\pi/2)] C_{ji}  \text{~ at }  i\in\alpha,
\end{equation}
For the channels $j\in\alpha$ this asymptotic expression has the
traditional sense, and the coefficients at the outgoing waves are
the elements of $S$-matrix, $C_{ji}=S_{ji}$, corresponding to the
transitions between the channels in the subspace $\alpha$. For the
channels $j\in\beta$ the first term in \eqref{eq13} is absent due
to $j\neq i$, whereas the second term ('outgoing' wave) is, in
fact, damped out:
 \begin{equation} \label{eq14}
\psi_{ji}(R) \sim \exp(-\mathrm{Im}k_j\cdot R) C_{ji}
      \text{~ at }  i \in\alpha, j \in\beta.
 \end{equation}
The density current of the outgoing  waves in these channels is
also damped out exponentially, therefore the coefficients $C_{ji}$
at $j\in\beta, i\in\alpha$ can not be treated as the $S$-matrix
elements of the transitions $(\alpha\to\beta)$.

It should be noted that the equation \eqref{eq13} is valid in the
far asymptotic region, where the centrifugal terms $L(L+1)/R^2$ in
the equations \eqref{eq11} can be neglected. Instead, we can
introduce the solutions in terms of the spherical Riccati-Hankel
functions
 \begin{equation} \label{eq15}
\psi_{ji}(R) \rightarrow k_i^{-1/2} h^{(-)}_{L_i}(k_i R)
 \delta_{ij} - k_j^{-1/2} h^{(+)}_{L_i}(k_i R) C_{ji} \quad (i\in\alpha,
 \text{ any } j).
\end{equation}
The spherical Riccati-Hankel functions are defined so that
\[ h^{(\pm)}_L(x)\rightarrow \exp[\pm\mathrm{i}(x-L\pi/2)] \]
at $x\rightarrow\infty$, therefore \eqref{eq15} at very large $R$
goes into \eqref{eq13}. But the solutions \eqref{eq15} are valid
also at the intermediate $R$ outside the interaction area as well.

Consider a possible scheme of the numerical solution of the system
\eqref{eq11} satisfying to the boundary conditions
$\psi_{ji}(0)=0$ at the origin as well as to the constraint
\eqref{eq12} at $i\in\beta$ and to the boundary conditions
\eqref{eq13} at $R\rightarrow \infty$. Let $N$ be the total number
of the channels included in the expansion \eqref{eq10}, $N_\alpha$
and $N_\beta$ the numbers of the channels in the subspaces
$\alpha$ and $\beta$, respectively. The total number of the
equations in the system \eqref{eq11} is also equal to $N$.
Therefore the system of the second-order differential equations
has $N$ linearly independent regular solutions $\mathbf{f}_p(R)$
with the components $f_{qp}(R)$, where the first index $q$ is the
number of the component, and the second index $p$ is the number of
the regular solution. In the practical calculations the different
regular solutions of the system are obtained, e.g., by using $N$
linearly independent initial values of the first derivatives
$\mathbf{f}'_p(0)$. Total set of the regular solutions can be
presented as the $(N\times N)$ matrix
$F(R)=\{\mathbf{f}_1(R),\mathbf{f}_2(R), \ldots,
\mathbf{f}_N(R)\}$.

Formally we could obtain the numerical solution for $F(R)$ from
the origin to the large $R$ and then find the linear combinations
of $\mathbf{f}_p(R)$ satisfying to the conditions \eqref{eq12} and
\eqref{eq13}. However this procedure leads to the essential
difficulties. The imaginary part of energy in the annihilation
channels produces the exponentially growing components of
$\mathbf{f}_p(R)$, which will be canceled in the correct linear
combination of $\mathbf{f}_p(R)$. But the remaining parts of the
functions can not be calculated with a good accuracy at
$R\rightarrow \infty$.

In order to avoid a loss of accuracy, we use the solution $F(R)$
only in the some restricted area $R\leq a$. For the area $R\geq a$
we construct the solutions of the system \eqref{eq11} beginning
from a some external point $R_e>R_0$, where the interaction
$V_{ji}(R)$ is negligible, and going to the smaller $R$ up to
$R=a$. The choice of the sewing point $a$ depends on the
parameters of the physical problem under consideration. It has to
be the subject of the numerical search in order to obtain a good
accuracy of the results.

Let $\mathbf{y}_k(R)$ be the solution of the system \eqref{eq11}
that contains at a large distance the single outgoing or damped
out wave in the channel $k$, so that
\begin{equation} \label{eq16}
y_{jk}(R)\rightarrow \delta_{jk} k_j^{-1/2} h^{(+)}_{L_j}(k_j R).
 \end{equation}
Total set of the $N$ linearly independent solutions forms the
$(N\times N)$ matrix $Y(R)$ that tends to the diagonal matrix at a
large $R$, but contains also non-diagonal elements in the
interaction area $(R\lesssim R_0)$. Similarly, we introduce the
$N_\alpha$ solutions $\mathbf{x}_i(R)$ that contain at a large $R$
the incoming waves in the channels $i\in\alpha$,
\begin{equation} \label{eq17}
x_{ji}(R)\rightarrow \delta_{ji} k_i^{-1/2} h^{(-)}_{L_i}(k_i R).
 \end{equation}
In line with the condition \eqref{eq12}, this set of the solutions
doesn't include the rising solutions at $i\in\beta$. We introduce
the square $(N\times N)$ matrix
\begin{equation} \label{eq18}
 X(R)= \begin{pmatrix} X_{\alpha\alpha} & 0 \\
                      X_{\beta\alpha} & 0 \end{pmatrix},
 \end{equation}
that contains the $N_\alpha$ columns of the $\mathbf{x}_i(R)$ and
the $N_\beta$ columns of zeros.

Total $(N\times N)$ matrix of the linearly independent solutions
of the system \eqref{eq11} satisfying to the condition
\eqref{eq12} and to the asymptotic boundary conditions
\eqref{eq12} can be presented in the form
\begin{equation} \label{eq19}
\Psi(R)= X(R)- Y(R) C,
\end{equation}
The matrix $\Psi(R)$ and the derivative $\Psi'(R)$ have to be
sewed at a point $R=a$ with a linear combination $F(R)A$ of the
regular solutions,
\begin{align}
F(a)A &= X(a)- Y(a) C, \label{eq20} \\
F'(a)A &= X'(a)- Y'(a) C. \label{eq21}
\end{align}
These equations allow to obtain the constant matrices $A$ and $C$.
Let $P(R)$ and $Q(R)$ be the $(N\times N)$ matrix solutions of the
system \eqref{eq11}. Introduce the matrix Wronskian
 \[W(P^T,Q)=P^T(R)Q'(R) - P'\,^T(R)Q(R),\]
where the superscript $T$ means the transposition of the matrix.
Due to the symmetry of the interaction in \eqref{eq11},
$V_{jk}(R)=V_{kj}(R)$, the Wronskian is a constant matrix, which
is independent on $R$. For the solutions $X(R)$ and $Y(R)$ we
obviously have
\begin{align}
W(X^T,X) &=W(Y^T,Y)=0, \label{eq22} \\
W(Y^T,X) &= -2 \mathrm{i} \begin{pmatrix} I_\alpha & 0 \\
                      0 & 0 \end{pmatrix} , \label{eq23}
\end{align}
where $I_\alpha$ is the unit matrix in the subspace $\alpha$.
Using the equations \eqref{eq20} and \eqref{eq21}, we get
\begin{align}
 A & = W^{-1}(Y^T,F) W(Y^T,X) , \label{eq24} \\
 C & = Y^{-1}(X-FA) =
  \begin{pmatrix} C_{\alpha\alpha} & 0 \\
                      C_{\beta\alpha} & 0 \end{pmatrix},
                      \label{eq25}
\end{align}
where all solutions $(X,Y,F)$ are taken in the sewing point $R=a$.

The submatrix $C_{\alpha\alpha}$, which gives the S-matrix of the
transitions in the subspace $\alpha$, can be expressed in terms of
the Wronskian matrices:
\begin{equation} \label{eq26}
S_{ji}\equiv C_{ji} = \left[W(X^T,F)W^{-1}(Y^T,F)\right]_{ji}
\quad \text{at } i,j \in \alpha.
\end{equation}
The S-matrix is not unitary, because the hamiltonian of the
problem is non-hermitian. The diagonal part of the unitary defect
gives the probability of the induced annihilation for the incoming
channel $i\in\alpha$,
\begin{equation} \label{eq27}
P(i\rightarrow annih)=1- \sum_{j\in\alpha}|S_{ji}|^2 .
\end{equation}
It allows to obtain the cross section of induced annihilation
\[ \sigma^a_i=(\pi/k_i^2)P(i\rightarrow annih). \]
Total cross section of the induced annihilation is a sum of
$\sigma^a_i$ with the relevant statistical weights of the
channels.

\section{Conclusion}

We have generalized the close-coupling method with annihilating
states included in the basis. The correct boundary conditions at
infinity suppose that the annihilating states are absent among the
incoming channels, but are presented among the outgoing channels
as the damped out waves. The S-matrix of the transitions in the
subspace $\alpha$ of all non-annihilating states is not unitary
due to the non-hermitian hamiltonian of the problem. The unitary
defect $(1- \sum |S_{ji}|^2)$ gives the cross section of induced
annihilation.

The scheme of the numerical solution of the system of
close-coupling equations is proposed. It contains the calculations
of two types of solution, from the origin to a point $R=a$ and
from a large $R$ to the point $a$, and the sewing of these
solutions at $R=a$. This scheme gives the S-matrix of the
transition between the ordinary channels as well as the
probability of the induced annihilation. The choice of the sewing
point $a$ depends on the parameters of the physical problem under
consideration.

Application of the developed approach to the problem of collisions
between antiprotonic helium ion $(\bar{p}\mathrm{He}^{2+})^*$ and
ordinary $\mathrm{He}$ atom at very low energy ($\sim 10$ Kelvin)
is given in the separate paper \cite{ref4}.

\section*{Acknowledgements}
We are grateful to V.N. Pomerantsev and V.P. Popov for useful
discussions. This work was supported by Russian Foundation for
Basic Research, Grant No. 06-02-17156.

\end{document}